\documentclass[aps,prl,twocolumn,amsmath,amssymb,showpacs]{revtex4}
\usepackage[english]{babel} \usepackage{latexsym}
\usepackage{graphicx} \usepackage{subfigure} %\usepackage{epsfig}
\usepackage{psfrag}

\usepackage{amssymb}
\usepackage{amsmath}
\usepackage{amscd}
\usepackage{here}
\usepackage{bbm}

\newcommand{\rr} {{\mathbf{r}}}
\newcommand{\pp} {{\mathbf{p}}}
\newcommand{\RCM} {{\mathbf{R}}}
\newcommand{\PCM} {{\mathbf{P}}}
\newcommand{\s} {{\mathbf{s}}}
\newcommand{\q} {{\mathbf{q}}}

\newcommand{\HH} {{\hat H}}

\begin{document}

\title{Bose-Fermi mixtures of self-assembled filaments of fermionic polar molecules}
\author{M.~Klawunn$^1$, J.~Duhme$^1$ and L.~Santos$^1$}
\affiliation{
\mbox{$^1$Institut f\"ur Theoretische Physik , Leibniz Universit\"at
Hannover, Appelstr. 2, D-30167, Hannover, Germany}}

\begin{abstract}
Fermionic polar molecules in deep 1D optical lattices may form self-assembled filaments 
when the electric dipoles are oriented along the lattice axis. These composites 
are bosons or fermions depending on the number of molecules per chain, leading to a 
peculiar and complex Bose-Fermi mixture, which we discuss in detail for the simplest
case of a three-well potential. We show that the interplay between filament binding energy, 
transverse filament modes, and trimer Fermi energy leads to a rich variety of possible 
scenarios ranging from a degenerate Fermi gas of 
trimers to a binary mixture of two different types of bosonic dimers.
We study the intriguing zero temperature and finite temperature physics of these composites 
for the particular case of an ideal filament gas loaded in 1D sites, and discuss possible methods to 
probe these chain mixtures.
\end{abstract}

\pacs{03.75.Lm,05.30.Jp} \maketitle

A new generation of experiments is starting to explore systems where
the dipole-dipole interaction (DDI) plays a significant and possibly dominant role.
Due to the long-range anisotropic character of the DDI, dipolar quantum gases offer a fascinating novel 
physics~\cite{Lahaye2009,Baranov2008}.
Exciting phenomena have been recently reported for experiments on 
Bose-Einstein condensates (BECs) of atomic magnetic dipoles, especially on
Chromium BEC~\cite{Chromium}, but also on spinor Rubidium BECs~\cite{Vengalattore2008},
Potassium~\cite{Fattori2008} and Lithium~\cite{Pollack2009}.
Magnetic atomic dipoles are however rather weak.
On the contrary heteronuclear molecules,
especially at their lowest rovibrational state,
may present a very large electric dipole moment
($\gtrsim 1$ Debye)~\cite{Ni2008, Ospelkaus2008, Deiglmayr2008}.
Although quantum degeneracy has not been yet achieved, 
the rapid pace of development allows to expect degenerate gases of polar molecules 
in the next future. These gases are expected to be largely dominated by the DDI.

Deep 1D optical lattices (formed by counter-propagating lasers) 
may slice a gas into non-overlapping samples. 
For non-dipolar (short-range interacting) particles these non-overlapping 
gases may be considered as independent parallel experiments.
The situation is completely different in dipolar gases, since the DDI leads to
 inter-site interactions. For weak DDI (e.g. atomic magnetic dipoles)
these inter-site interactions lead to scattering between BECs 
at different sites~\cite{Nath2007}, collective excitations 
shared by non-overlapping sites~\cite{Klawunn2008}, 
or damping of Bloch oscillations~\cite{Fattori2008}.
For bosonic polar molecules the non-local dipolar effects 
may be much stronger, leading to 
fascinating effects as pair-superfluidity for ladder-like
lattices~\cite{Arguelles2007} and filament BEC~\cite{Wang2006}.

Filamentation is indeed an interesting possibility introduced by the DDI. 
This phenomenon, first suggested in the context of ferrofluids by de Gennes and Pincus~\cite{deGennes1970}, 
has attracted a considerable theoretical interest for the case of classical dipoles~\cite{Teixeira2000}. 
Dipolar chains in classical ferrofluids were recently observed in superparamagnetic iron colloids~\cite{Butter2003} 
and single-domain magnetite colloids~\cite{Klokkenburg2006}. In Ref.~\cite{Wang2006} it was shown that 
a similar filamentation process may occur for bosonic polar molecules in deep lattices, which 
may organize into chains sustained by an attractive inter-site DDI, forming a dipolar chains liquid (DCL) which may Bose-condense~\cite{Wang2006}.

%figure1
\begin{figure}%[ht]
\vspace{0.5cm}
\begin{center}
\includegraphics[height=1.7cm,width=4.5cm]{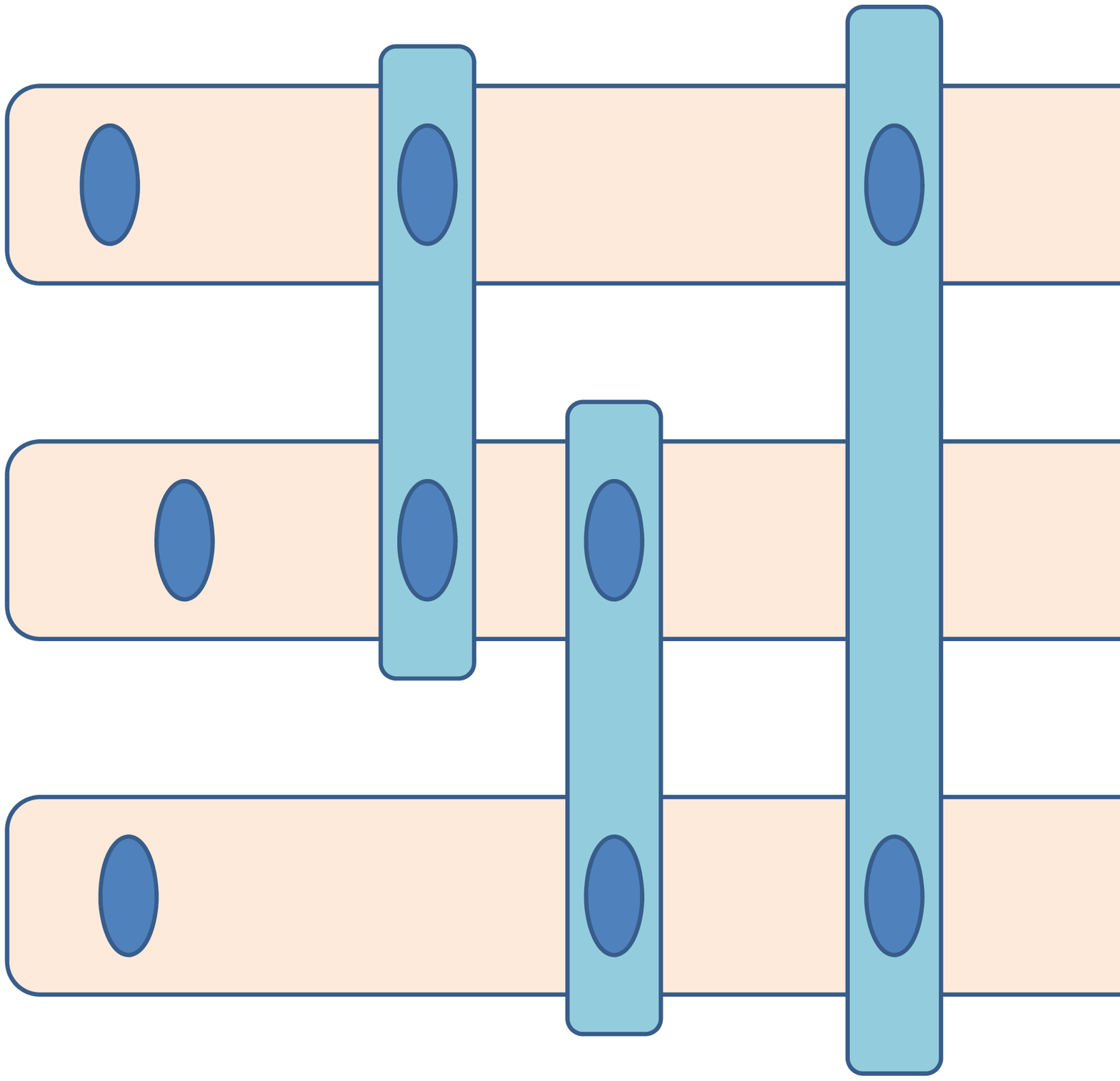}
\end{center}
%\vspace*{-0.5cm}
\caption{Polar fermionic molecules in a three-well potential may remain unpaired, form fermionic trimers, 
or bosonic dimers between nearest neighbors or next-nearest neighbors.}
\vspace*{-0.2cm}
\label{fig:1}
\end{figure}

%Fermionic filaments
In this Letter we consider DCLs of fermionic polar molecules. 
Far from being a trivial extension of the bosonic case, fermionic polar molecules lead to a 
very different and rich physics. Whereas for bosonic molecules the chains are obviously bosons, 
for fermionic molecules the bosonic or fermionic character of the filaments depends on the number of molecules 
in the chain. This has particularly relevant consequences when the number of available lattice sites is odd.
Here we focus on the simplest non-trivial case, namely a three-well potential (Fig.~\ref{fig:1}).
For simplicity we restrict our discussion to the ideal gas regime, where inter-filament interactions are neglected. 
Although this approximation is of limited quantitative validity (and would demand mesoscopic samples in specific 
1D arrangements as discussed below), 
it contains already many of the qualitatively new features which may be expected for more general scenarios of polar Fermi molecules 
in deep 1D and 2D optical lattices. In particular, the competition between trimer/dimer binding
and trimer Fermi energy results in a non-trivial dependence of the character of the DCL as a function of 
the number of molecules per site $N$. If $N$ is smaller than a critical $N_c$ the DCL is a Fermi-degenerate gas of 
trimers. However for $N>N_c$ the trimers coexists with a Bose mixture formed by  
pseudo-spin-1/2 dimers and spinless dimers, leading to a peculiar Bose-Fermi mixture. We show that these 
Bose-Fermi DCLs may be probed by monitoring the spatial distribution of the molecules.

%The system under consideration}

We consider fermionic polar molecules with mass $m$ and electric dipole $d$ 
in a deep three-well potential along the $z$-direction, with inter-site spacing $\Delta$. 
This arrangement may be created 
by e.g. selectively emptying all sites of a strong 1D optical lattice except 
three neighboring ones. The potential barriers are large-enough to prevent any 
inter-site hopping. Due to reasons discussed below, the analysis of the problem simplifies 
notably if the gas is considered as strongly confined along the $y$-direction (e.g. 
in a single node of a lattice as that in the $z$-direction). Along the remaining $x$-direction 
we consider a shallow harmonic confinement with frequency $\omega$. 
The molecules interact via the DDI
$V_d(\mathbf{r})= d^2 (1-3\cos^2\theta)/r^3$, where $\theta$ is the angle formed by $\mathbf{r}$ with the dipole orientation.  We assume that the dipoles form an angle $\alpha$ with the $z$ direction, such that $\sin^2\alpha=1/3$. 
Although this particular orientation and the 1D character of the sites are not needed for the 
formation of the DCL gas, which may occur also in stacks of 2D sites~\cite{Wang2006}, this particular scenario 
allows both for a strong attraction between dipoles placed on top of each other and for a vanishing DDI between molecules at the same site. This largely reduces the inter-filament interaction, allowing for a simplified 
ideal gas scenario, as discussed below.

The attraction between polar molecules placed on top of each other may be strong-enough 
to bind two or more polar molecules into self-assembled chains (Fig.~\ref{fig:1}). 
Whereas for bosonic molecules these chains are in any case bosons~\cite{Wang2006}, 
for fermionic molecules the fermionic/bosonic character of the filaments depends on the odd/even number of 
molecules in a given chain. In particular, the three-well configuration 
allows for fermionic trimers (and of course monomers), and two different kinds of bosonic dimers, namely 
those between two molecules at nearest neighbors (type I dimers), and those between two molecules at 
the uppest and lowest site (type II dimers) (Fig.~\ref{fig:1}). Note that dimers I are actually 
pseudo-spin-$1/2$ bosons, since dimers in sites $1$ and $2$ are not equivalent to dimers in sites $2$ and $3$.

%single filament physics

The ground-state of a single filament 
of $M$ molecules is calculated as for bosonic molecules~\cite{Wang2006}, and hence 
we just sketch for completeness the basic ideas. 
Let $\rr_j$ ($\hat\pp_j$) be the position (momentum) operator of a molecule at site $j$. 
Introducing $\hat\PCM=\sum_{j=1}^M\hat\pp_j$, $\RCM=\sum_{j=1}^M\rr_j/M$, $\q_j=\pp_j-\PCM$, 
$\s_j=\rr_j-\RCM=\{x_j,y_j,z_j\}$, the Hamiltonian splits into $\HH=\HH_{CM}+\HH_{rel}$, 
where $\HH_{CM}=\hat\PCM^2/2Mm+Mm\omega^2 R_x^2/2$ describes 
the filament center-of-mass and
\begin{eqnarray}
\HH_{rel}&=&\sum_{j=1}^M \left [
\frac{\hat\q_j^2}{2m}+
\frac{m}{2}\left ( \omega_\perp^2 (y^2+z_j^2)+\omega^2 x^2\right ) 
\right ] \nonumber \\
&+&\sum_{i,j>i} V_{d}\left [ \s_i-\s_j \right ].
\end{eqnarray}
the relative motion. The on-site $yz$ confinement is approximated by a strong isotropic harmonic 
oscillator of frequency $\omega_\perp$.  The wavefunction of the $j$-th molecule is chosen as 
$\psi_j(x_j-x_{j0})\varphi_j(y_j)\varphi_j(z_j-z_{j0})$, where 
$\varphi_j(\eta)=\exp \left ( -\eta^2/2l_\perp^2 \right ) / \sqrt{l_\perp\sqrt{\pi}}$, and 
$\psi_j(\eta)=\exp \left ( -\eta^2/2R_0^2 \right ) / \sqrt{\sqrt{\pi} R_0}$, with $l_\perp=\sqrt{\hbar/m\omega_\perp}$ 
and $R_0$ is the variational width of the $x$ wavepackets. For deep lattices one may 
approximate $l_\perp\rightarrow 0$ (energy corrections are $\lesssim 1\%$ for
depths $>14$ recoil energies $\hbar^2\pi^2/2m\Delta^2$). 

\begin{figure}%[ht]
\begin{center}
\includegraphics[width=0.32\textwidth]{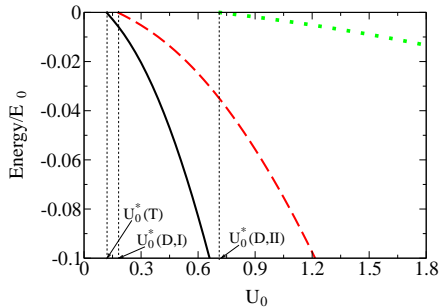}
\end{center}
\vspace*{-0.5cm}
\caption{Binding energy (in units of $E_0=\hbar^2/m\Delta^2$) of the different 
composites of Fig.~\ref{fig:1} as a function of $U_0=md^2/\hbar^2\Delta$.}
\vspace*{-0.2cm}
\label{fig:2}
\end{figure}

Minimizing the energy of straight filaments 
($x_{j0}=0$) with respect to $R_0$ we obtain the filament binding energy 
\footnote{For simplicity we neglect that the minimal-energy configuration is slightly tilted from the vertical 
with an angle $\sim \pi/10$.}.
We denote as $-E_T$, $-E_{D,I}$ and $-E_{D,II}$ the binding energies 
for, respectively, trimers, dimers I, and dimers II. These energies grow 
with the dipole strength $U_0=md^2/\hbar^2\Delta$. There exists a critical $U_0^*$ 
such that for $U_0<U_0^*$ the composites unbind ($R_0\gtrsim l_{HO}=\sqrt{\hbar/m\omega}$). 
Note that $U_0^*(T)<U_0^*(D,I)\ll U_0^*(D,II)$ (Fig.~\ref{fig:2}) due to the different strength of the DDI 
in each composite. In the following we consider the regime $U_0> U_0^*(D,II)$, where 
$R_0\ll l_{HO}$ for all of the possible composites of Fig.~\ref{fig:1}.

Transverse filament excitations contribute to the gas entropy, being relevant at finite temperature $T$. 
In addition, and contrary to the case of bosonic molecules~\cite{Wang2006}, transverse modes are important 
for fermionic molecules also at very low $T$ since they may significantly reduce the trimer Fermi energy. 
For a chain of $M$ molecules, we obtain 
the low-lying modes $\xi_{\nu=1,\dots,M}$ after 
expanding the chain energy $E$ around its minimum, and diagonalizing  $\partial^2E/\partial x_j\partial x_{j'}$, where $j,j'=1,\dots,M$.

% Quantum statistics

In the following we consider the filament statistics, assuming an ideal filament gas.
This largely simplifies the analysis of the problem, while 
allowing for the discussion of key qualitative features of these systems, in particular the competition between 
different Bose and Fermi composites. 
This approximation is just quantitatively valid for mesoscopic samples in the arrangement discussed above, 
as we discuss at the end of this Letter.

The fermionic or bosonic character of the chains is reflected by the 
average occupations for trimers, dimers I, dimers II and monomers:
\begin{eqnarray}
N_T(n,\nu_T)&=&\left [ e^{\beta[-E_T+\xi_{\nu_T}+\epsilon_n-(2\mu_1+\mu_2)]}+1 \right ] ^{-1} \label{NT}\\
N_{D,I}(n,\nu_{D,I})&=&\left [ e^{\beta[-E_{D,I}+\xi_{\nu_{D,I}}+\epsilon_n-(\mu_1+\mu_2)]}-1 \right ]^{-1} \label{NDI}\\
N_{D,II}(n,\nu_{D,II})&=&\left [ e^{\beta[-E_{D,II}+\xi_{\nu_{D,II}}+\epsilon_n-2\mu_1]}-1 \right ]^{-1} \label{NDII}\\
N_{S,j}(n)&=&\left [ e^{\beta[\epsilon_n-\mu_j]}+1 \right ]^{-1} \label{NS}
\end{eqnarray}
where $N_{S,j}$ denotes the average occupation of individual molecules at site $j$, $\xi_{\nu_{T;D,I;D,II}}$ the 
transverse filaments modes of the different composites, $\epsilon_n=\hbar\omega(n+1/2)$ the harmonic oscillator levels and 
$\beta=1/k_B T$ the inverse temperature. 
In the previous expressions we have assumed symmetric configurations such that the number of 
dimers I in sites $1$--$2$ is the same as the number of dimers I in sites $2$--$3$, and equal to $N_{D,I}(n,\nu_{D,I})$.
Note that $\mu_1=\mu_3$ is the chemical potential for molecules at the uppest and lowest sites, whereas 
$\mu_2$ denotes the chemical potential for molecules in the middle site. These different chemical potentials are necessary 
to fulfill the normalization conditions, in which we assume $N$ molecules per lattice site. Imposing 
symmetry between the uppest and the lowest sites, these conditions acquire the form:
\begin{eqnarray}
N&=&N_T+N_{D,I}+N_{D,II}+N_{S,1}, \label{Norm1}\\
N&=&N_T+2N_{D,I}+N_{S,2}, \label{Norm2}
\end{eqnarray}
where $N_T$, $N_{D,I}$, $N_{D,II}$, $N_{S,1}$ and $N_{S,2}$ denote respectively the total number of trios, dimers I in sites 
$1$--$2$ (or $2$--$3$), dimers II, monomers in site $1$ (or $3$) and monomers in site $2$. From (\ref{Norm1}) and (\ref{Norm2}) 
we obtain $\mu_1(N,T)$ and $\mu_2(N,T)$, and from (\ref{NT}--\ref{NS}) the occupation 
numbers.

Due to the attractive DDI between molecules in the filament, the most bound chain is the trimer. 
The difference in binding between dimers and trimers induces that 
for sufficiently small $N$ and at
low-enough T the DCL becomes a degenerate Fermi gas of trimers. The trimers  
fill up oscillator levels (and also transverse trimer modes) up
to the corresponding Fermi energy $E_F(N)$, which equals $N\hbar\omega$ for rigid filaments 
but it is actually smaller due to the transverse trimer modes.
However, if the number of molecules per
site is sufficiently large, the growth in Fermi energy overcomes the binding energy difference. 
This transition may be easily estimated by comparing the average energy per molecule
for the case of two trimers and that for the case of $2$ dimers I and one dimer II. 
This leads to a condition for the critical number of molecules per site $N_c(U_0,\omega)$,
$E_F(N_c)=2E_T-3(E_{D,I}+E_{D,II})/2$ (which we have confirmed numerically). Note that $N_c$ grows with growing 
$U_0$ and decreasing $\omega$. For $N<N_c$ the DCL is a degenerate trimer gas, whereas for $N>N_c$ 
the trimer gas coexists with a mixture of pseudo-spin-$1/2$ bosons (dimers I) and spin-less bosons (dimers II). 

%figure2

% Spatial distribution

\begin{figure}%[ht]
\begin{center}
\includegraphics[width=3.5cm,height=7.0cm,angle=270]{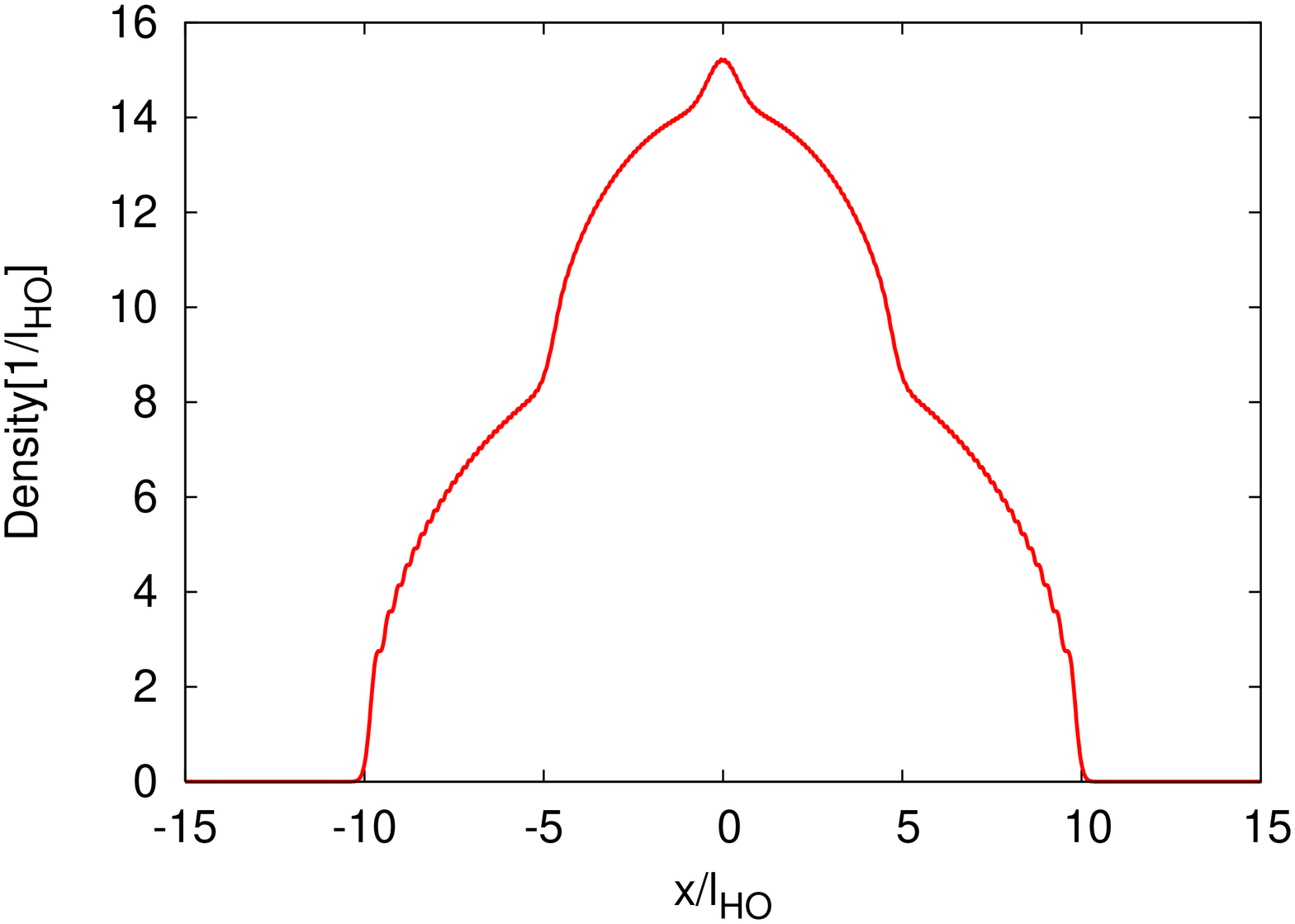}
\includegraphics[width=3.5cm,height=7.0cm,angle=270]{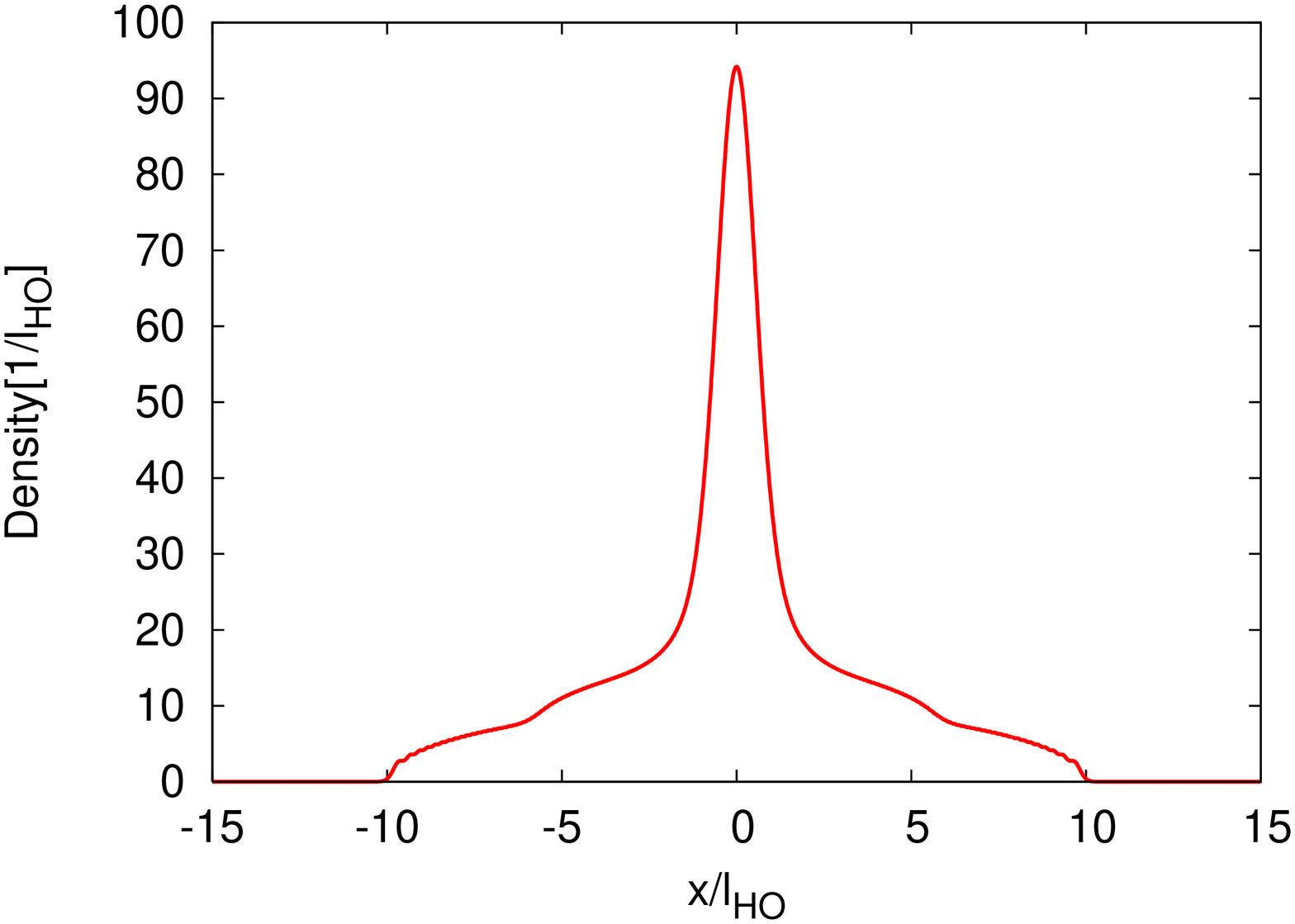}
\end{center}
\vspace*{-0.5cm}
\caption{Integrated density profiles of the molecules, for  
$\xi_{1_T}/\omega<N<N_c$ (top), and $N>N_c$ (bottom). We consider $U_0=2$, $\omega/2\pi=1$Hz, $m=100$amu, which lead to $N_c=230$.} 
\vspace*{-0.2cm}
\label{fig:3}
\end{figure}

The peculiar properties of the DCL translate into the spatial molecular distribution 
integrated over the three sites. For $N<N_c$ and $N<\xi_{1_T}/\omega$, only trimers in their 
internal ground state are formed, and hence the gas behaves as a spin-less Fermi gas of particles of mass $3m$, 
presenting a Thomas-Fermi density profile $~(1-(x/R)^2)^{1/2}$ 
with $R/l_{HO}=\sqrt{2N/3}$. For $\xi_{1_T}/\omega<N<N_c$, the DCL is still a trimer gas, but 
transversal modes may be populated. In that case the density profile departs from the Thomas-Fermi profile (Fig.~\ref{fig:3}, top), due 
to the appearance of internally excited trimers in low harmonic oscillator levels. For $N>N_c$ the density profile changes dramatically. 
Note that since we consider 1D gases, dimer BEC is strictly speaking precluded. 
However, due to finite size the dimers quasi-condense (at low-enough $T$) 
occupying the few lowest levels of the harmonic oscillator. Hence when $N$ surpasses $N_c$ 
a Bose cloud nucleates at the trap center. As a result the distribution of the polar molecules shows a 
Gaussian-like peak at the trap center (Fig.~\ref{fig:3}, bottom).

%figure3

% Temperature

\begin{figure}%[ht]
\begin{center}
\includegraphics[width=0.32\textwidth]{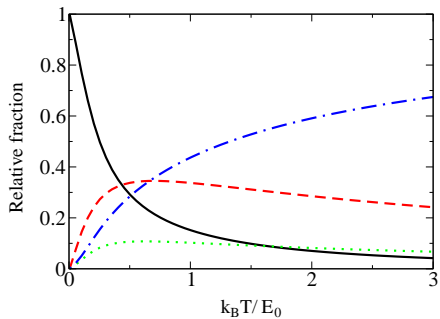}
\end{center}
\vspace*{-0.5cm}
\caption{Temperature dependence of the fraction of molecules in 
trimers (solid), dimers I (dashed), dimers II (dotted) and monomers (dashed-dotted). We consider the parameters of Fig.~\ref{fig:3} with $N<N_c$.} 
\vspace*{-0.2cm}
\label{fig:4}
\end{figure}

The DCL presents as well an intriguing finite temperature physics due to the role of filament 
modes and the different binding energy of dimers and trimers. This is particularly clear from a finite $T$ 
analysis of a DCL with $N<N_c$ (Fig.~\ref{fig:4}).
Note that whereas at very low $T$ the DCL is purely a trimer Fermi gas, at finite $T$  
it becomes more favorable to populate dimers than to populate higher excited trimer states. As a consequence 
the system presents a striking thermal enhancement of the bosonic
modes. Interestingly, contrary to the standard situation, this leads to a maximal 
central peak density for a given finite $T$. For even larger $T$ the central density decreases again due to the 
occupation of dimers at higher oscillator modes, and the breaking of the filaments into individual molecules. 

%figure4

% N>>Nc

For $N\gg N_c$ and $U_0>U_0^*(D,II)$ the DCL is at low T  a basically 
pure Bose gas of dimers I and II (except for a small trimer fraction). 
Since both dimers have mass $2m$, the difference between them cannot be 
discerned from the analysis of the integrated density profile of the molecules. However the different binding energy and excited 
dimer modes for both types of dimers may be studied spectroscopically to reveal the dual nature of the mixture. 
If $N\gg N_c$ but $U_0^*(D,I)\ll U_0<U_0^*(D,II)$, dimers II are precluded, and hence the DCL will become at low $T$ a 
Bose-Fermi mixture of dimer-I bosons and degenerate monomers at sites $1$ and $3$ (which act as a pseudo-spin-$1/2$ fermions). Again, this exotic mixture could be revealed from the corresponding dual density profile.

% Interactions

The discussed ideal gas analysis allows for a relatively easy understanding of key qualitative features 
characterizing fermionic polar molecules in deep optical lattices under more general conditions, 
as the competition between filament-binding energy and Fermi energy of fermionic chains, 
the relevant role of the filament modes at zero and finite $T$, 
or the formation of peculiar mixtures of different types of composite bosons and fermions. 
However, the quantitative validity of the ideal gas approximation is rather limited (also 
for bosonic molecules~\cite{Wang2006}), even for the previously discussed 
1D arrangement with the particular choice for the dipole orientation. We may estimate the importance of the 
inter-filament interactions by comparing the inter-trimer interactions with the Fermi energy of rigid trimers ($\epsilon_F=N\hbar\omega$).
For deeply bound chains ($R_0\sim \Delta/2$) and at inter-filament distances $x> \Delta$ 
we may approximate the interaction between molecules at different chains 
as that between two point dipoles. Adding up these interactions we may estimate the 
mean inter-trimer DDI $V_{ff}(\bar{x})$, where $\bar{x}$ is the mean inter-trimer distance.
For the case of $d=0.8$ Debye, $m=100$ atomic mass units, $\Delta=0.5\mu$m, and $\omega/2\pi=1$Hz, we obtain 
$U_0\simeq 2$, and $N_c\simeq 230$. For this value $\bar{x}\simeq 1.7\Delta$ and $V_{ff}\simeq 0.33\epsilon_F$. 
The ideal gas approximation is hence quantitatively valid only for dilute mesoscopic samples (as those 
considered in our numerical calculations). 
Once the dimer Bose gas nucleates at the trap center the ideal gas condition is quickly violated, 
due to the larger bosonic densities, although the formation of the dual density profile (similar to that in Fig.~\ref{fig:3}) 
still holds. For stacks of 2D sites the ideal gas approximation fails even for extremely low 2D densities. 
However, the formation of dimer mixtures beyond a given critical density should also occur 
for 2D arrangements. These mixtures may be considered as weakly-interacting for 2D densities $n$ such that 
$nr_*^2<1$ with $r_*=md^2/\hbar^2$. For the previous values this demands $n\lesssim 1.1\times 10^8$cm$^{-2}$. 
For $N\gg N_c$ (and $U_0>U_0^*(D,II)$) weakly-interacting dimers will form a BEC of three 
different bosons (dimers I in $1$--$2$, I in $2$--$3$, and II), whose properties will largely depend 
on the precise determination of the different inter-dimer interactions, 
which will be the subject of a future work.

% Conclusions

Summarizing, fermionic polar molecules in three-well potentials are expected to form a rather peculiar 
filament gas. Depending on the filling per site and the interaction strength we expect that the 
character of the chain gas ranges from a pure trimer gas at low fillings, to a bosonic mixture of pseudo-spin-$1/2$ and spin-less 
dimers for large-enough fillings and dipole strengths. Note, finally, that molecules in even larger number of sites 
may form a quantum gas mixture of increasing complexity. Dipolar chain liquids are hence an exciting perspective 
for on-going experiments with polar fermionic molecules.

\acknowledgments

We thank G. Shlyapnikov for useful discussions. This work was supported by the DFG 
(SFB407, QUEST), and the ESF (EUROQUASAR).


\begin{thebibliography}{99}

\bibitem{Lahaye2009} Th. Lahaye et al., arXiv:0905.0386.

\bibitem{Baranov2008} M. A. Baranov,
Physics Reports 2008 {\bf 464}, 71.

\bibitem{Chromium} A. Griesmaier {\it et al.},
Phys. Rev. Lett. {\bf 94}, 160401 (2005);
J.~Stuhler {\it et al.}
Phys. Rev. Lett. {\bf 95} 150406 (2005);
Q.~Beaufils {\it et al.},
Phys. Rev. A {\bf 77}, 061601(R) (2008);
Th.~Lahaye {\it et al.}
Nature {\bf 448}, 672 (2007).

\bibitem{Vengalattore2008} M. Vengalattore {\it et al.},
Phys. Rev. Lett. {\bf 100}, 170403 (2008).

\bibitem{Fattori2008} M. Fattori {\it et al.},
Phys. Rev. Lett. {\bf 101}, 190405 (2008).

\bibitem{Pollack2009} S. E. Pollack {\it et al.},
Phys. Rev. Lett. {\bf 102}, 090402 (2009).

\bibitem{Ni2008} K. K. Ni {\it et al.},
Science {\bf 322}, 231 (2008)

\bibitem{Ospelkaus2008} S.Ospelkaus {\it et al.},
Nature Phys. {\bf 4}, 622 (2008).

\bibitem{Deiglmayr2008} J. Deiglmayr {\it et al.},
Phys. Rev. Lett. {\bf 101}, 133004 (2008).

\bibitem{Nath2007} R. Nath, P. Pedri, and L. Santos,
Phys. Rev. A {\bf 76}, 013606 (2007).

\bibitem{Klawunn2008} M. Klawunn and L. Santos, arXiv:0812.3543

\bibitem{Arguelles2007} A. Arg\"uelles and L. Santos,
Phys. Rev. A {\bf 75}, 053613 (2007).

\bibitem{Wang2006} D. Wang, M. D. Lukin, E. Demler,
Phys. Rev. Lett. {\bf 97}, 180413 (2006).

\bibitem{deGennes1970} P. G. de Gennes and P. A. Pincus, Phys. Kondens.Mater. 11, 189 (1970).

\bibitem{Teixeira2000} For a review see e.g. P. I. C. Teixeira, J. M. Tavares and M. M. Telo da Gama, J. Phys. Condens.Matter 12, R411 (2000).

\bibitem{Butter2003} K. Butter {\it et al.}, Nat. Mater. {\bf 2}, 88 (2003).

\bibitem{Klokkenburg2006} M. Klokkenburg {\it et al.}, Phys. Rev. Lett. {\bf 96}, 037203 (2006).


\end{thebibliography}
\end{document}